\documentclass{ITaSconf}
\usepackage[utf8x]{inputenc}
\usepackage[english]{babel}
\usepackage{url}
\usepackage{amsmath}
\usepackage{amssymb}
\usepackage{mathrsfs}
\usepackage{graphicx}
\usepackage{epstopdf}

\usepackage[ruled]{algorithm}
\usepackage[noend]{algorithmic}
\newcommand{\T}{^{\text{\tiny\sffamily\upshape\mdseries T}}}

\newcommand\mylim[1]{\mathop{\operator@font #1}\limits}

\title{Inverse Protein Folding Problem\\ via Quadratic Programming}
\author{
  Andrii Riazanov \\
  \begin{affiliation}
	Skolkovo Institute of Science and Technology\\
    Moscow Institute of Physics and Technology\\
  \end{affiliation}\\
  \email{andrei.ryazanov@phystech.edu}
  \and
Mikhail Karasikov\\
  \begin{affiliation}
    Moscow Institute of Physics and Technology\\
    Skolkovo Institute of Science and Technology\\
  \end{affiliation}\\
  \email{karasikov@phystech.edu}
  \and 
  Sergei Grudinin \\
\begin{affiliation}
University of Grenoble Alpes, LJK\\
CNRS, LJK\\ Inria
\end{affiliation}\\
\email{sergei.grudinin@inria.fr}
}

\begin{document}
\maketitle
\begin{abstract}
This paper presents a method of reconstruction a primary structure of a protein that folds into a given geometrical shape. This method predicts the primary structure of a protein and restores its linear sequence of amino acids in the polypeptide chain using the tertiary structure of a molecule.
 Unknown amino acids are determined according to the principle of energy minimization. This study represents inverse folding problem as a quadratic optimization problem and uses different relaxation techniques to reduce it to the problem of convex optimizations. Computational experiment compares the quality of these approaches on real protein structures.
\end{abstract}
\section{Introduction}
A protein is a sequence of basic molecules of amino acids. Proteins are essential compounds of all living organisms, they are involved in almost all structural, catalytic, sensory, and regulatory functions. Properties and functions of proteins mostly depend on their structure. Protein engineering is aimed at studying of protein structures and it finds applications in a number of areas. For example, the design of proteins with desired shapes and properties is required in such areas as medicine, biotechnology, synthetic biology, nanotechnologies, etc.\par

	While {\em computational protein folding} predicts the structure of a given sequence, the goal of {\it computational inverse protein folding problem} is to find amino acid sequences that would fold best into a given 3D shape or scaffold. It can be referred as {\it computational protein design} (CPD), which targets at choosing a sequence of amino acids to perform a particular task. As there are 20 possible amino acids for each position in the protein chain, the variety of sequences for a specific 3D shape grows exponentially. In CPD, the inverse folding problem is formulated as an optimization problem, aimed at the minimization of the energy of the protein.\par
	
	\par

	The inverse folding problem is usually solved including the fact that each residue can adopt multiple conformational states, which are most often grouped into a finite number of conformational isomers (rotamers). Two approximations are common for this problem. First, it is assumed that the protein's backbone is fixed. Second, the continuous domain of available conformations for the side-chains is approximated using a finite set of discrete conformations, defined by inner dihedral angles. The CPD is then formulated as the problem of determining the conformation corresponding to the  minimum energy from that finite set. \par
	 
A variety of algorithms was developed to solve the inverse folding problem. In \cite{CPD}, the problem is expressed as a Cost Function Network (CFN), also known as the Weighted Constraint Satisfaction Problem (WCSP). The interaction energies are represented as cost functions, and the unknown variables are the sets of pairs {\it (amino acid, conformation)}. Then, the problem is solved using the dead-end elimination$/A^*$ algorithm ($DEE/A^*$). More precisely, first, the $DEE$ algorithm reduces the computational cost of the problem, excluding from the consideration conformations that are unlikely to be optimal. Then, the $A^*$ algorithm allows to expand a sequence-conformation tree, so that sequence-conformations are extracted and sorted on the basis of their energy values. The CFN problem is solved using the {\it toulbar2} solver. The investigation \cite{CPD} also compares the CFN approach with several other optimization models. CPD is expressed as a  0/1 linear programming (01LP) problem, which is defined by a linear criterion to optimize over a set of Boolean variables under a conjunction of linear equalities and inequalities. It is also represented as a problem of 0/1 quadratic programming (01QP), which has a quadratic criterion to optimize with the same types of constraints as in 01LP. To solve these both models, the authors used the {\it cplex} solver and compared the results with the CFN approach. \par

In \cite{GA}, the inverse folding problem is solved using genetic algorithm. This algorithm optimizes the sequence of amino acids in a protein chain with a stochastic search similar to natural selection in evolution. This approach was later improved in \cite{MO} by expressing the problem as a highly multi-modal optimization problem. This is achieved by using a diversity measure as the objective function through multi-objectivization. The common feature of all genetic algorithms is that they rely on evolutionary information of existing structures.

Investigation \cite{ASP} studies the Answer Set Programming (ASP) approach for the inverse folding problem. It uses $DEE$ in combination with branch and bound algorithms to eliminate high number of non-optimal pairs {\it (amino acid, conformation)}, as in \cite{CPD}. Then the problem is represented in the ASP model and solved with solver {\it clingo4}.\par
All the aforementioned investigations determined both amino acid and conformation for all positions in the polypeptide chain. These approaches used energy functions which consider interactions between all possible pairs of rotamers, for example, energy function, implemented in the CPD dedicated tool {\it osprey 2.0} \cite{osprey}. The mentioned works compare the efficiency of different algorithms, i.e. the time these algorithms need to find the minimum of energy for given instances. The CFN model demonstrated the best results overall, although even it could not solve all the instances in reasonable time. However, the quality of prediction of the primary structure (the coincidence between the found structure and the native one) is not investigated in these works. This quality depends on the choice of energy potential, and there is a challenging problem to find a practical scoring potential, which does not always match with the true potential function \cite{POT}.\par
	
The main distinction of our investigation is that only the amino acids in the protein chain are predicted, and spatial conformations (rotamers) are not considered. This approach decrease the computational complexity of the problem significantly. It allows to receive basic information about the primary structure of a protein using much less resources, and allows to find the sequence of amino acids even for the most complicated instances. Thus, this study uses energy functions \cite{POT2, DFIRE}, which describe only interactions between all possible pairs of amino acids, but not rotamers.\par

In this study the inverse protein folding problem is reduced to the quadratical programming problem, which can be written as
\begin{equation}
\label{1}
\begin{split}
	&\text{minimize}\ \ \ f_0(\vec{x})  \\ 
	&\text{subject\ to}\ \ f_i(\vec{x}) \le 0,\quad i = 1, 2, \dots, N,
\end{split}
\end{equation}
where \[ f_i (x) = \vec{x}\T\mathbf{A}_i\vec{x} + 2\vec{b}\T_i\vec{x} + c_i,  i = 0, 1, \dots, N.\] 
The initial problem is not convex, because the matrices $\mathbf{A}_i$ are generally indefinite, therefore the problem is NP-hard and can not be solved efficiently. This paper studies semidefinite relaxation 
, Lagrangian relaxation \cite{REL} and continuous relaxation of the problem~\eqref{1} to convex one, which can be solved in practical time. The solution of the relaxed problem provides a lower bound of the optimal value of~\eqref{1}. We use the rounding of the relaxed solution to find an upper bound and an approximate the optimal value. 
This paper also suggests Sequential Quadratic Programming (SQP) \cite{SQP} and Simulated Annealing 
approaches to solve \eqref{1}. Finally, we compare these relaxations and techniques and estimate the efficiency and the quality of primary structure prediction of using different approaches on real proteins structures from the SCWRL4 test set \cite{SCW}.

\section{Problem statement}
Let the protein chain consist of $N$ amino acids. The set $\mathcal{C} = \{1, 2, \dots, 20\}$ contains indexes that encode all~20~possible amino acids. Let $\vec{y} = (y_1,\ldots,y_N)$ be the sequence of residues, where $y_i \in \mathcal{C}$. Let functions $E_{kl}:\;\mathcal{C}^2\to\mathbb{R}$ define symmetrical pairwise interaction energies between residues $k$ and $l$. Then the problem of energy minimization is represented as
\begin{equation}   
\label{2}
	\sum_{j=1}^N\sum_{i=1}^NE_{ij}(y_i,y_j) \rightarrow \underset{y_1,y_2,\dots,y_N \in \mathcal{C}}{\min}.
\end{equation}
The problem can also be written in the following way:
\begin{equation}
\label{3}
\begin{aligned}
	&\underset{\vec{x}=[\vec{x}_1\T,\dots,\vec{x}_N\T]\T}{\text{minimize}} &&\vec{x}\T\mathbf{Q}\vec{x} \\
	&\text{subject to} &&\vec{x}_k \in \{0,1\}^{20}, \quad &k=1,\dots, N,		\\
	&		         &&\|{\vec{x}_k}\|_0 =1,\quad &k=1,\dots,N,
\end{aligned}
\end{equation}
where

\[ \mathbf{Q} = 
\begin{bmatrix} [E_{11}] & [E_{12}]& \cdots & [E_{1N}] \\
		     [E_{21}] & [E_{22}]& \cdots & [E_{2N}] \\
		     \vdots	& \vdots	&	\ddots      &   \vdots \\
		     [E_{N1}] & [E_{N2}]& \cdots & [E_{NN}] \\
\end{bmatrix}, 						\]
\[E_{ij} = 
\begin{bmatrix} E_{ij}(c_1,c_1) & E_{ij}(c_1,c_2)& \cdots & E_{ij}(c_1,c_{20}) \\
		     E_{ij}(c_2,c_1) & E_{ij}(c_2,c_2)& \cdots & E_{ij}(c_2,c_{20}) \\
		     \vdots	& \vdots	&	\ddots      &   \vdots \\
		     E_{ij}(c_{20},c_1) & E_{ij}(c_{20},c_2)& \cdots & E_{ij}(c_{20},c_{20}) \\
\end{bmatrix}, 						\]
\begin{center}
 $\mathbf{Q} \in \mathbb{R}^{20N\times20N},\quad E_{ij} \in \mathbb{R}^{20\times20} $.
\end{center}\par\bigskip
Finally, it is practical to re-write the problem as
\begin{equation}
\label{4}
\begin{aligned}
	&\underset{\vec{x}\in\{0,1\}^{20N}}{\text{minimize}} &&\vec{x}\T\mathbf{Q}\vec{x} \\
	&\text{subject to} && \mathbf{A}\vec{x} = \mathbf{1}_N,
\end{aligned}
\end{equation}
where 
\[ \mathbf{A} = 
\begin{bmatrix} 1 \cdots 1 & 0 \cdots 0 & \cdots\cdots & 0\cdots 0 \\
		     0 \cdots 0 & 1 \cdots 1 & \cdots\cdots & 0\cdots 0 \\
		     \vdots	& \vdots	&	\ddots      &   \vdots \\
		     \underbrace{0 \cdots 0 }_{20}& \underbrace{0 \cdots 0 }_{20} & \cdots\cdots & \underbrace{1 \cdots 1 }_{20} \\   
\end{bmatrix},  \quad\mathbf{A} \in \{0,1\}^{N\times20N}. \]\par

To estimate the quality of primary structure prediction, we use the scoring matrix BLOSUM62 \cite{BLOS}. 
This matrix is a substitution matrix used for calculating the degree of matching between two protein sequences.

Let $\mathbf{B}$ represents the BLOSUM62 matrix. Then, the score $\mathbf{B}(y_i, \hat{y}_j)$ is the score for the substitution of residue $y_i$ for $\hat{y}_j$ at $j$-position in the chain. If the score is positive, the residues are interchangeable (or equal). Otherwise, this substitution is less likely to occur, and the prediction is wrong. Let $\vec{y}_\text{nat}, \vec{y}_\text{pred} \in \mathcal{C}^N$ be native and predicted sequences of length $N$, respectively. Then, the quality function can be defined as
\begin{equation}
\label{QUAL}
S(\vec{y}_\text{nat}, \vec{y}_\text{pred}) = \dfrac{\sum_{k= 1}^{N} \mathbf{B}((\vec{y}_\text{nat})_k, (\vec{y}_\text{pred})_k)}{\sum_{k= 1}^{N} \mathbf{B}((\vec{y}_\text{nat})_k, (\vec{y}_\text{nat})_k)} 
\end{equation}
A good quality prediction corresponds to $S > 0$, and the best predictions have values of $S$ close to $1$. Negative values of $S$ mean poor quality predictions.\par
The aim of this study is to compare different optimization approaches to solve the problem~\eqref{4}. The value of $S$ is considered as the quality of the primary structure prediction, while the optimization power of algorithms is estimated from the ability of the algorithm to predict the sequence with the lowest possible energy.

\section{Algorithms and relaxations} \label{meth}
\subsection{Greedy algorithm}
Given a problem \eqref{2}
\[	\sum_{j=1}^N\sum_{i=1}^NE_{ij}(y_i,y_j) \rightarrow \underset{y_1,y_2,\dots,y_N \in \mathcal{C}}{\min}, \] consider the greedy search algorithm to find an approximate solution:\par
\begin{algorithm}[h]
\caption{Greedy algorithm}
\label{greed}
	\algorithmicrequire{$\ \vec{y}_\text{start} = (y^0_1,\dots,y^0_N),\ y^0_k \in \mathcal{C},\ k=\overline{1,N}$ }

	\algorithmicensure{$\ \vec{y}_\text{pred} = (y_1,\dots,y_N)$}
\begin{algorithmic}
	\REPEAT
	\STATE{
		$num\_changed\gets 0$
	
		\FORALL{$k \in \{1,\dots,N\}$}
			\STATE{
				$\hat{y}_k:=\mathop{arg\,min}\limits_{y'_k\in\mathcal{C}}f(y_1,\dots,y_{k-1},y'_k,y_{k+1},\dots,y_N)$\\
				\IF{$\hat{y}_k\neq y_k$}
					\STATE{$y_k:=\hat{y_k}$}
					\STATE{$num\_changed:=num\_changed+1$}
				\ENDIF
			}
		\ENDFOR
	}
	\UNTIL{$num\_changed > 0$}
\end{algorithmic}
\end{algorithm}
On each iteration, this algorithm searches for a protein conformation of a lower energy than the current one among conformations in the neighborhood of the current conformation. On each step it determines for each position of the chain whether there exists such an amino acid $\hat{y}_k$ that would decrease the energy of the protein by substitution $y_k$ with $\hat{y}_k$. It stops when there is no position to change, so this algorithm returns at least the local minimum of the problem \eqref{2}.

\subsection{Continuous relaxation}
The matrix $\mathbf{Q}$ in \eqref{4} could be relaxed to semidefinite positive by shifting its spectrum. Denote 
\begin{equation}
\label{66}
\hat{\mathbf{Q}} = \mathbf{Q} - \lambda_\text{min}(\mathbf{Q}) \cdot \mathbf{I}_{20N}, \quad c = \lambda_\text{min}(\mathbf{Q})\cdot N.
\end{equation} 
Then
\[ \vec{x}\T\hat{\mathbf{Q}}\vec{x} = \vec{x}\T\left(\mathbf{Q} - \lambda_\text{min} \mathbf{I}_{20N}\right)\vec{x} =
 \vec{x}\T\mathbf{Q}\vec{x} - \lambda_\text{min}\vec{x}\T\vec{x}.\]
 Considering $\vec{x} \in \{0, 1\}^{20N}$ and $\mathbf{A}\vec{x} = \mathbf{1}_N$, we have $\vec{x}\T\vec{x} = N$. Therefore, optimization problem~\eqref{4} can be written as:
 \begin{equation}
\label{6}
\begin{aligned}
	&\underset{\vec{x}\in\{0,1\}^{20N}}{\text{minimize}} &&\vec{x}\T\hat{\mathbf{Q}}\vec{x} + c\\
	&\text{subject to} && \mathbf{A}\vec{x} = \mathbf{1}_N.
\end{aligned}
\end{equation}
Now the matrix $\hat{\mathbf{Q}}$ is semidefinite positive due to \eqref{66}, because all its eigenvalues are nonnegative. 
Consider continuous relaxation of \eqref{6} by relaxing constraints $\vec{x} \in  \{0, 1\}^{20N}$ to $\vec{x} \ge \vec{0}_{20N}$ :

 \begin{equation}
\label{7}
\begin{aligned}
	&\underset{\vec{x}\ge \vec{0}_{20N}}{\text{minimize}} &&\vec{x}\T\hat{\mathbf{Q}}\vec{x} + c\\
	&\text{subject to} && \mathbf{A}\vec{x} = \mathbf{1}_N.
\end{aligned}
\end{equation}
 The problem \eqref{7} is convex now, and therefore can be solved efficiently. The solution of \eqref{7} also gives the lower bound for the optimum of the primary problem \eqref{4}. An approximate solution of \eqref{4} and an approximate optimum are found from $\vec{x}=[\vec{x_1}\T,\dots,\vec{x_N}\T]\T$ as follows: 
 \[ y_k = \underset{j =\overline{1,20}} {arg\,max}\  x_k^j, \quad k = 1,\dots,N. \]
 The solution could be also improved by using the greedy algorithm \ref{greed} with the solution of \eqref{7} as a starting position. The final approximation will be the upper bound for the optimum value of the initial problem \eqref{4}.

 \subsection{Semidefinite relaxation (SDP)}
 SDP relaxation is described in \cite{SDP2}. Applying it to the problem \eqref{4} we have
\begin{equation}
\notag
\begin{aligned}
& \underset{\substack{ \ \ \mathbf{X}\in\mathbb{R}^{20N{\times}20N}\\ \vec{x} \ge \vec{0}_{20N}}}{\text{minimize}} & & \text{Tr}\left(\mathbf{Q}\mathbf{X}\right) \\
& \text{subject to} & & \mathbf{X}\succeq\vec{x}\vec{x}\T, \\
&			& & X_{ii} = x_i,\ i=\overline{1,20N}, \\
&			& & \mathbf{A}\vec{x} = \vec{1}_N,
\end{aligned}
\end{equation}
where $\mathcal{S}_{+}^n=\{\mathbf{X}\in\mathbf{R}^n:\ \mathbf{X}=\mathbf{X}\T\succeq 0\}$ is the set of symmetric semidefinite positive matrices of dimension~$N$.
By Schur's complement \cite{BV_CVX}, the problem is equivalent to
\begin{equation}
\label{SDP}
\begin{aligned}
& \underset{\substack{ \ \ \mathbf{X}\in\mathbb{R}^{20N{\times}20N}\\ \vec{x} \ge \vec{0}_{20N}}}{\text{minimize}}
& & \text{Tr}\left(\mathbf{Q}\mathbf{X}\right) \\
&\text{subject to} && 
\begin{bmatrix}
	\mathbf{X} & \vec{x}\\
	\vec{x}\T & 1
	\end{bmatrix}
	\in\mathcal{S}_{+}^{20N+1}, \\
&&& X_{ii} = x_i,\ i=\overline{1,20N}, \\
&&& \mathbf{A}\vec{x} = \vec{1}_N.
\end{aligned}
\end{equation}
Again, the relaxed problem is convex and can be solved efficiently using the methods of convex programming. The optimum of \eqref{SDP} gives the lower bound for \eqref{4}. The binary vector $\vec{x}$ and the upper bound for the initial problem \eqref{4} are recovered from the solution in the same way as in the previous case (continuous relaxation).

\subsection{Lagrangian relaxation}
The Lagrangian relaxation \cite{REL, BV_CVX} is a good method for getting a cheaply computable lower bound on the optimal value of a non-convex quadratic optimization problem. This method uses the fact that the dual of a problem is always convex, and the weak duality guarantees the optimum value of a dual problem to be the lower bound to the initial one. \par
The Lagrangian of~\eqref{4}~is
\begin{equation}
\notag
\begin{aligned}
L(\vec{x}, \vec{\lambda}, \vec{u})
&= \vec{x}\T \mathbf{Q} \vec{x} + \sum\limits_{i = 1}^{20N}\lambda_i(x_i^2 - x_i) + \vec{u}\T (\mathbf{A}\vec{x} - \vec{1}_N) = \\
&= \vec{x}\T \left(\mathbf{Q} + \mathbf{D}(\vec{\lambda})\right) \vec{x} - \vec{\lambda}\T \vec{x} + \vec{u}\T (\mathbf{A}\vec{x} - \vec{1}_N) = \\
&= \vec{x}\T \left(\underbrace{\mathbf{Q} + \mathbf{D}(\vec{\lambda})}_{\mathbf{P}(\vec{\lambda})}\right) \vec{x}
    - \left(\underbrace{\vec{\lambda}\T - \vec{u}\T \mathbf{A}}_{\vec{q}\T(\vec{\lambda},\vec{u})}\right) \vec{x}
    - \underbrace{\vec{u}\T \vec{1}_N}_{r(\vec{u})}.
\end{aligned}
\end{equation}
The dual function is then
\begin{equation}
\notag
\begin{aligned}
&g(\vec{\lambda}, \vec{u}) = \inf_{\vec{x}\in\mathbb{R}^{20N}} L(\vec{x}, \vec{\lambda}, \vec{u})
= \inf_{\vec{x}\in\mathbb{R}^{20N}} \vec{x}\T \mathbf{P}(\vec{\lambda}) \vec{x} - \vec{q}\T(\vec{\lambda},\vec{u})\vec{x} - r(\vec{u}) = \\
&=	\begin{cases}
    -\frac{1}{4} \vec{q}\T(\vec{\lambda},\vec{u}) \mathbf{P}^{+}(\vec{\lambda}) \vec{q}(\vec{\lambda},\vec{u}) - r(\vec{u}), & \text{if } \mathbf{P}(\vec{\lambda})\succeq 0 \text{ and }\\ &\vec{q}(\vec{\lambda},\vec{u}) \perp \text{Ker}\,\mathbf{P}(\vec{\lambda}),\\
    -\infty, \quad \text{otherwise}.
  \end{cases}
\end{aligned}
\end{equation}
Finally, the Lagrangian relaxation of \eqref{4} is:
\begin{equation}
\label{lagr}
\begin{aligned}
& \underset{\vec{\lambda}\in\mathbb{R}^{20N},\,\vec{u}\in\mathbb{R}^N}{\text{maximize}}
& & \gamma - r(\vec{u})\\
&\text{subject to} &&
\gamma \leq 0, \\
&&& \begin{bmatrix}
	\mathbf{P}(\vec{\lambda}) & \frac{1}{2}\vec{q}(\vec{\lambda},\vec{u})\\
	\frac{1}{2}\vec{q}\T(\vec{\lambda},\vec{u}) & -\gamma
	\end{bmatrix}
	\in\mathcal{S}_{+}^{20N+1}. \\
\end{aligned}
\end{equation}

 \subsection{Sequential Quadratic Programming}
Sequential Quadratic Programming (SQP) \cite{SQP, SQPBook} is one of the most effective methods for constrained nonlinear optimization. The idea of SQP is to solve nonlinear problem using a sequence of quadratic programming (QP) subproblems. The objective function of the subproblem on each step is a convex quadratic approximation of the Lagrangian function.\par
First, we relax the constraints of the initial problem \eqref{4} in order to apply the SQP method. The constraints $x_i \in \{0,1\}$ should be substituted for $0\le x_i\le 1$: 
\begin{equation}
\label{cont}
\begin{aligned}
	&\underset{\vec{x}\in \mathbb{R}^{20N}}{\text{minimize}} &&\vec{x}\T\mathbf{Q}\vec{x} \\
	&\text{subject to} && \mathbf{A}\vec{x} = \mathbf{1}_N,\\
	&		          && \mathbf{0}_{20N} \le \vec{x} \le \mathbf{1}_{20N}  
\end{aligned}
\end{equation}
The Lagrangian of \eqref{cont} is
\begin{equation}
\label{cont_lagr}
\begin{aligned}
L(\vec{x}, \vec{\lambda}, \vec{\mu}, \vec{u})
&= \vec{x}\T \mathbf{Q} \vec{x} - \vec{\lambda}\T\vec{x} + \vec{\mu}\T(\vec{x} - \mathbf{1}_{20N}) + \vec{u}\T (\mathbf{A}\vec{x} - \vec{1}_N).
\end{aligned}
\end{equation}
Now let us formulate the QP subproblem for the $k$-th step of the algorithm:
\begin{equation}
\label{QP}
\begin{aligned}
	&\underset{\vec{d}\in\mathbb{R}^{20N}}{\text{minimize}} &&\dfrac1{2}\vec{d}\T\mathbf{H}_k\vec{d} + 2\vec{x_k}\T\mathbf{Q}\vec{d} \\
	&\text{subject to} && \mathbf{A}\left(\vec{x_k+d}\right) = \mathbf{1}_N,\\
	&		          && \mathbf{0}_{20N} \le \vec{x_k+d} \le \mathbf{1}_{20N}.  
\end{aligned}
\end{equation}
Here $\mathbf{H}_k$ is a positive definite approximation of the Hessian matrix of the Lagrangian function \eqref{cont_lagr}, and $\vec{x_k}$ is a current iterate. The solution of \eqref{QP} is used to form a new iterate $(\vec{x}_{k+1}, \vec{\lambda}_{k+1}, \vec{\mu}_{k+1}, \vec{u}_{k+1})$.\\
Each iteration of the algorithm consists of three steps:
\begin{enumerate}
\item Updating the Hessian Matrix
\item Solving a Quadratic Program
\item Line Search and Merit Function
\end{enumerate}
1. At each iteration a quasi-Newton approximation of the Hessian $\nabla^2_{xx}L(\vec{x}_{k+1}, \vec{\lambda}_{k+1}, \vec{\mu}_{k+1}, \vec{u}_{k+1})$ is calculated using the Broyden-Fletcher-Goldfarb-Shanno (BFGS) \cite{BFGS} method. Denote
\[ \vec{s}_k = \vec{x}_{k+1} - \vec{x}_k, \]\[\qquad \vec{y}_k = \nabla_xL(\vec{x}_{k+1} \vec{\lambda}_{k+1}, \vec{\mu}_{k+1}, \vec{u}_{k+1}) - \nabla_xL(\vec{x}_k, \vec{\lambda}_{k}, \vec{\mu}_{k}, \vec{u}_{k})\]
Then the BFGS approximation for the next iteration has the form
\begin{equation}
\mathbf{H}_{k+1} = \mathbf{H}_{k} + \dfrac{\vec{y}_k\vec{y}_k\T}{\vec{y}_k\T\vec{s}_k} - \dfrac{\mathbf{H}_k\vec{s}_k\vec{s}_k\T\mathbf{H}_k\T}{\vec{s}_k\T\mathbf{H}_k\vec{s}_k}.
\end{equation}
This approximation keeps $\mathbf{H}_k$ positive definite during the iterations. 
\\[0.3cm]
2. With iterate $(\vec{x}_k, \vec{\lambda}_{k}, \vec{\mu}_{k}, \vec{u}_{k})$ and a quasi-Newton approximation of Hessian $\mathbf{H}_k$, the subproblem QP \eqref{QP} is formulated. Considering the condition on $\mathbf{H}_k$ to be positive definite, QP is convex and can be solved efficiently. New iterates for dual variables $\vec{\lambda}_{k+1}, \vec{\mu}_{k+1}, \vec{u}_{k+1}$ are determined from the solution immediately as the values of the dual variables of \eqref{QP} in the optimal point. The new iterate for the primal variable is set to $\vec{x}_{k+1} = \vec{x}_k + \alpha_k\vec{d}_k$, where $\vec{d}_k$ is the solution of QP \eqref{QP}, and the optimum length of the step $\alpha_k$ is determined at the next step.\\[0.3cm]
3. The step length parameter $\alpha_k$ is determined in order to produce a sufficient decrease in a merit function. We use the following function: 
\begin{equation}
\label{merit}
\Psi(\vec{x}) = \vec{x}\T\mathbf{Q}\vec{x}  + \vec{w}\T\vec{x}_{+} + \vec{p}\T\left(\vec{\mathbf{1}_{20N}-\vec{x}}\right)_++ \vec{r}\T\left(\mathbf{A}\vec{x}-\mathbf{1}_N\right),
\end{equation}
Here, if $\vec{m} = \left[ \vec{w}\T, \vec{p}\T, \vec{r}\T \right]\T$, then $\vec{m}$ is calculated as
\[ \vec{m}_i = (\vec{m}_{k+1})_i= \max\left\{ \pi_i, \dfrac{(\vec{m}_k)_i +\pi_i}{2}\right\}, \quad i=1,\dots, 41N,\]
where vector $\vec{\pi} = \left[ \vec{\lambda}\T, \vec{\mu}\T, \vec{u}\T\right]\T$ is the vector of dual variables of QP \eqref{QP}.
The parameter $\alpha_k$ is calculated then from the minimization of the univariate function
\[\Psi(\vec{x}_k + \alpha_k\vec{d}_k) \rightarrow \underset{\alpha_k}{\text{min}}.\]\par\bigskip
The problem \eqref{cont} is then reduced to a sequential solving of convex quadratical optimization problems.
%

 \subsection{Simulated Annealing}
Simulated Annealing (SA) \cite{SA2} is used to compare different approaches. It is a probabilistic method, and although it is unlikely to find the global optimum solution, it may often find a solution very close to it.\par
Let $T : \mathbb{Z}_+ \rightarrow (0; +\infty)$ be the exponentially decreasing function of the temperature, and $\mathbf{S}$ be a finite set of {\bf states} -- all manner of possible sequences. The set $N(x)$ is a set of {\bf neighbors} to the state $x \in \mathbf{S}$. The energy function $F(x)$ is the objective function. $x_0 \in \mathbf{S}$ is the starting position. Then the iteration of SA algorithm is:
\begin{enumerate}
\item Randomly choose the state from the neighbors of a current state $x_{k+1} \in N(x_k)$
\item Assess the energy of the chosen state $F_{k+1} = F(x_{k+1})$
\item Compare $F_{k+1}$ and $F_k$ and decide, whether to accept selected move. If $F_{k+1} < F_{k}$, the move is accepted. Else, the move is accepted with the probability $\exp{\left(-\dfrac{F_{k+1}-F_k}{T(k)}\right)}.$
\end{enumerate}\par
In this investigation problem, the set $\mathbf{S}$ is the set of all possible sequences of amino acids for a given length, neighbors $N(x)$ for the sequence $x$ are the sequences, in which 5 or less residues differ from $x$. The energy function $F(x)$ is the energy of the protein structure.

\section{Solution}\par
This section describes the process of solving the problem \eqref{4} using methods, presented in the previous section.
%
First, we build the energy matrix for a protein structure. In the current study, we use two coarse-grained distance-dependent potentials. One of them is described in \cite{POT2}. Another is DFIRE-$C_a$ potential \cite{DFIRE}, whose parameters are kindly provided by the authors. They both require only the distance between two $C_{\alpha}$ carbon atoms in a pair of amino acids to estimate the energy of interaction between these residues. So, by extracting coordinates of $C_{\alpha}$ atoms of the backbone of a given protein, we can build the energy matrix. That is, for each pair of positions $i, j$ in the chain, we calculate the distance $d_{ij}$ between $C_{\alpha}$ atoms at these positions. Then, the energy of the interaction between all $210$ pairs of amino acids located at distance $d_{ij}$ are estimated, and this is how matrices $E_{ij}$ are obtained. By doing so for all $i, j \in \overline{1, N}$, we build the matrix $\mathbf{Q}$.\par
Then different relaxations are applied to the problem \eqref{4}:
\begin{itemize}
\item Lagrangian relaxation \eqref{lagr} is used to receive the lower bound for the optimum value. It does not provide an approximate solution to the problem, because there is no strong duality between \eqref{4} and its dual.
\item Greedy algorithm \ref{greed} is initialized with a random starting position.
\item Continuous and SDP relaxations find both lower bound and approximate optimum of the problem. We improve received solutions with the greedy algorithm in the next step.
\item SQP and Simulated Annealing give an approximate optimum of the energy.
\end{itemize}
Finally, we  estimate the quality of optimization and prediction for each method at each test protein structure. The quality of prediction is obtained from \eqref{QUAL} using the BLOSUM62 matrix.

\section{Computational details}

All the tested algorithms were implemented in {\it Python}, using the CVXPY package \cite{cvxpy} to solve different convex problems that are obtained after the proposed relaxations. For the SQP algorithm, we used its MATLAB implementation from the Global Optimization Toolbox. The computation experiment was set on a quad-core Intel  Core(TM) i7-4700HQ CPU @ 2.40GHz PC with 12 GB of RAM running on Ubuntu 14.04 LTS OS.

\section{Results and Discussion}
To compare the quality of different relaxation approaches in solving non-convex quadratic problems, we performed a series of computational experiments. More precisely, we used six algorithms, described in the section \ref{meth},  to find the minimum of the protein energy.
The computational experiments were set on a test set of protein structures extracted from the SCWRL4 benchmark.
The average length of sequences for these proteins was 110 amino acids, so the dimension of $\vec{x}$ was about 2000. 

\begin{figure}[t]
\includegraphics[width=1.1\linewidth]{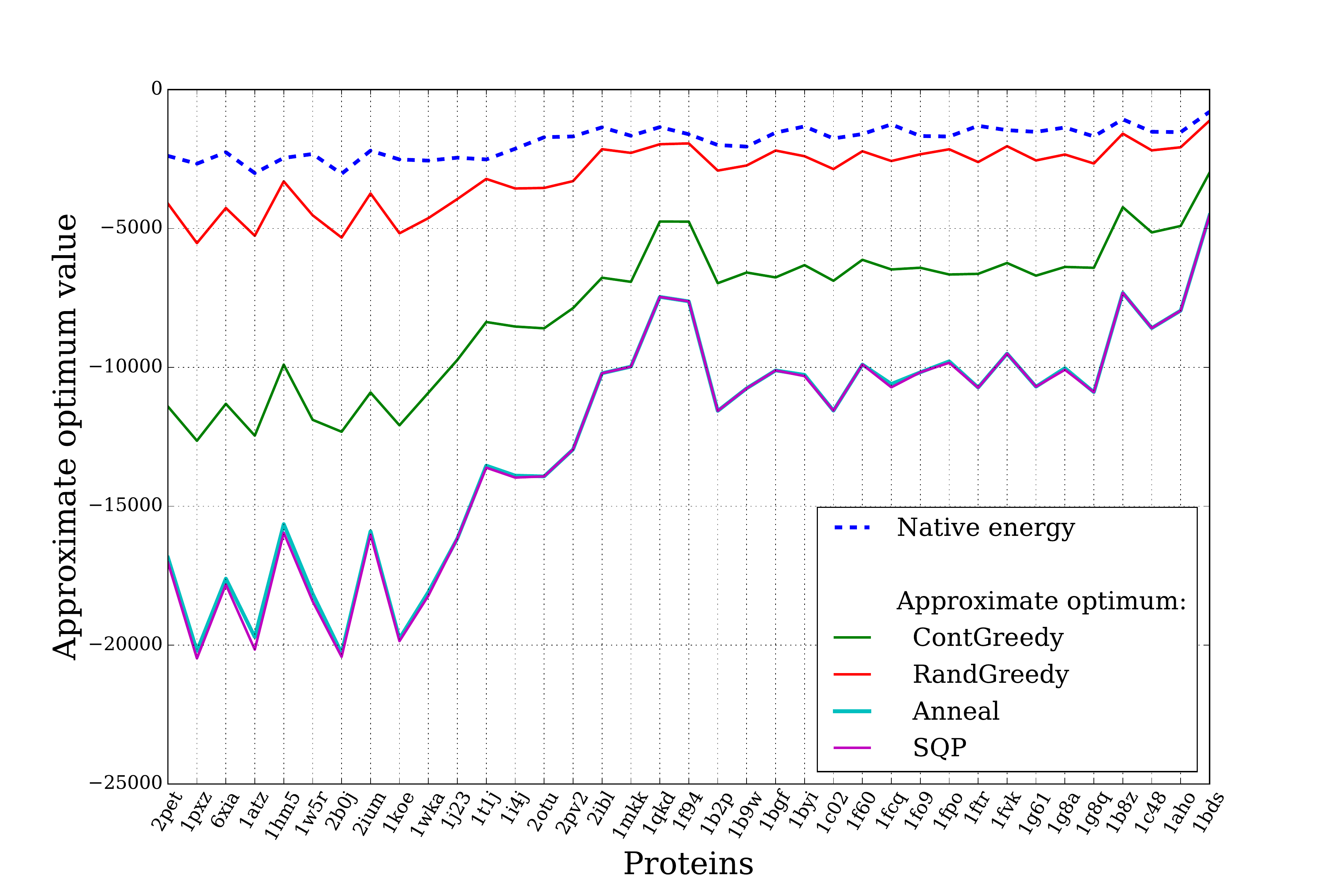}
\caption{Approximate energy optimum for different relaxations computed on the test set.}
\label{Res}
\end{figure}

Figure \ref{Res} presents the results of optimization the energy using the potential from \cite{POT2}.

From this figure we can see that among the compared methods, SQP and SA demonstrated the best power of optimization. They returned very proximate optimum energy  values and predicted similar sequences of amino acids. The continuous relaxation approach minimizes the objective function worse than SA and SQP, but still returns the sequence with the energy, lower than the native. We should note that the SDP and Lagrangian relaxations methods cannot be applied to large optimization problems. This is because the number of their variables is quadratic with respect to the length of vector $\vec{x}$, and also because the computational requirements for these approaches exceed the available resources for the sequences of a practical length. For example,  for $N=100$ the number of variables for these methods is more than $2\cdot 10^6$.

\begin{figure}[t]
\includegraphics[width=1.1\linewidth]{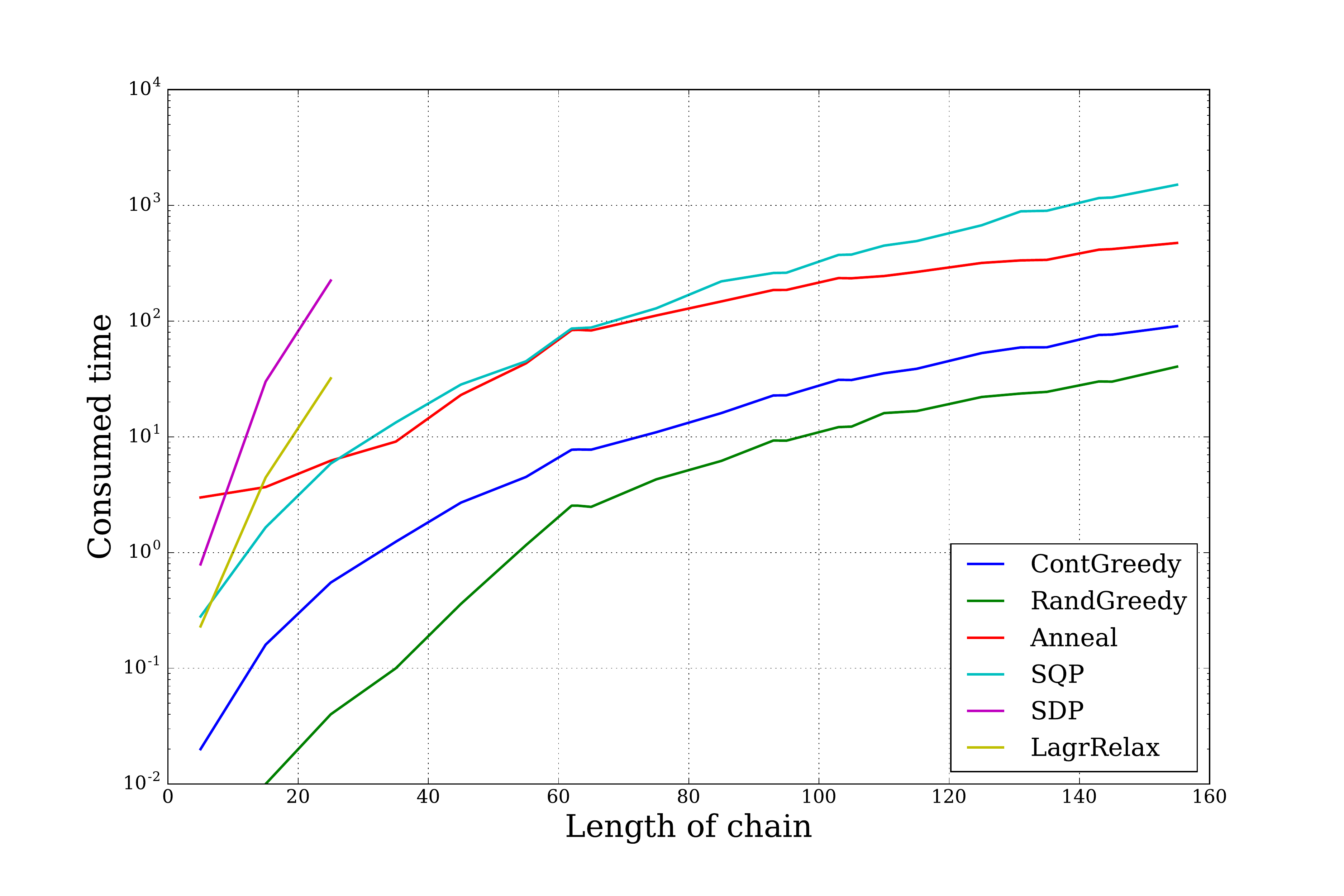}
\caption{Computational time as a function of the length of a protein chain for all the algorithms in logarithmic scale.}
\label{time_log}
\end{figure}

\begin{figure}[t]
\includegraphics[width=1.1\linewidth]{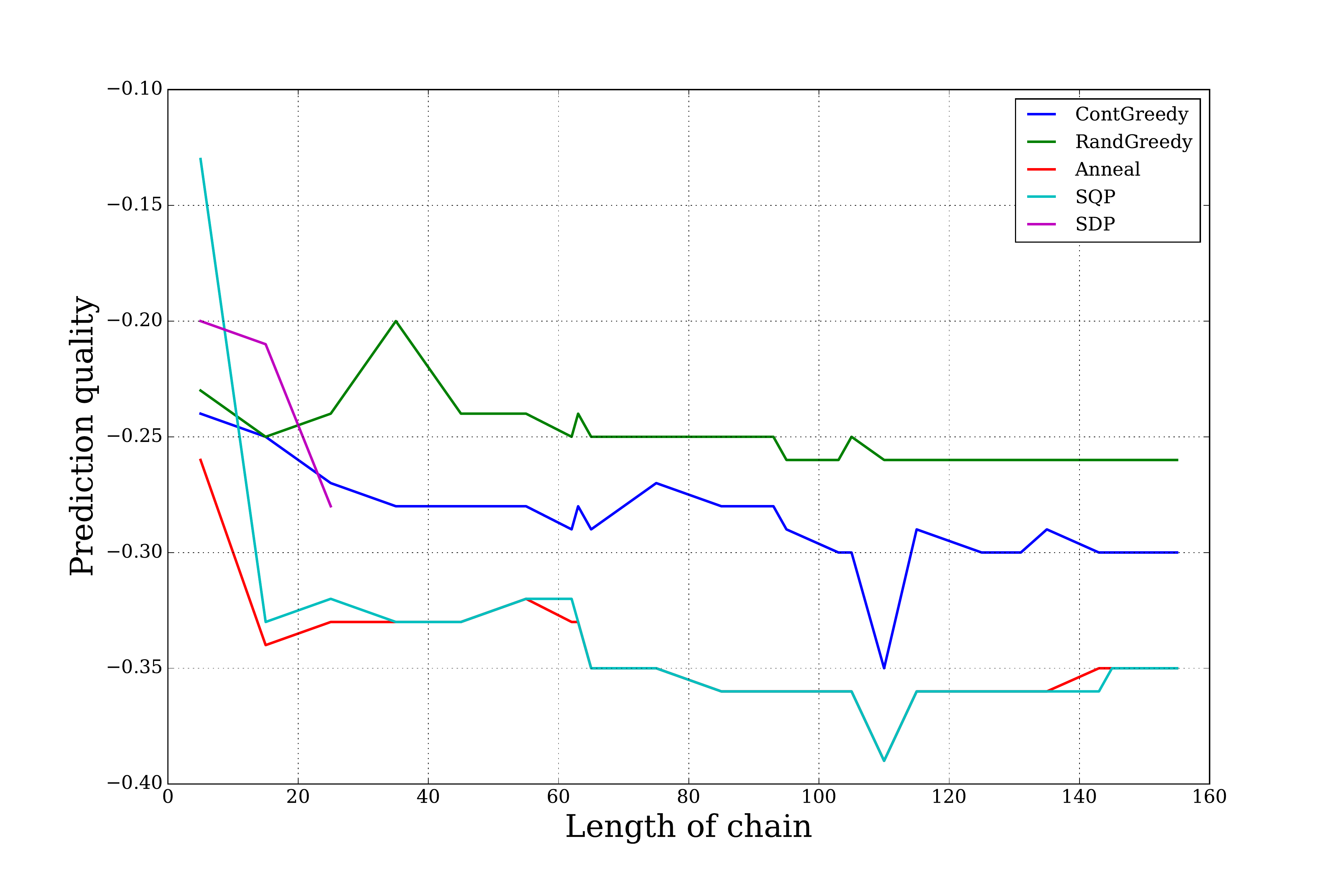}
\caption{Dependence of prediction quality on the length of a protein chain for all the tested methods. }
\label{quality}
\end{figure}

For all the techniques and all the test structures, we measured the time spent on finding the solution of the optimization problem. It was done in the following way -- for different lengths of sequences from 5 to 155, we truncated the length of the structures of proteins this upper limit. Then, we applied all the tested optimization methods  to solve the problem with cropped backbones, and the average time for each length was measured. 
Also, we computed the mean of quality of prediction for such sequences. The dependences of computational time and quality on the length of a protein chain are shown in Figs. \ref{time_log} and \ref{quality}, respectively.

Figure \ref{Res} clearly demonstrates that all the relaxations and all the tested methods found sequences of amino acids with the energy lower, than the one of the native state. The greedy algorithm with a native sequence of amino acids as the starting position also returned the sequence with a lower energy for all the tested potentials. This means that the native structures are not even the local minimum for the potential energy function used in this experiment. Therefore, the two tested potentials do not satisfy the crucial assumption, i.e., that the native conformation has the lowest energy between all possible states for a given fixed backbone, and thus the quality of primary structure prediction is extremely poor. 
In searching for a minimum of energy, the algorithms select residue types with the lowest contact energies. This happens to be observed for pairs of cysteine-cysteine and tyrosine-tyrosine. As the result, the predicted structure often consist fully either of cysteine residues, or of tyrosine residues.

\section{Conclusion}\par
In this paper we proposed to formulate the Inverse Protein Folding Problem as a quadratic optimization problem, where the objective function is the energy of the protein, which is assumed to have the lowest value for the native conformation.
The results shows that the SQP and SA approaches have the best optimization power on the test set of protein structures extracted from the SCWRL data set. These two methods returned very similar approximate optimums for nearly all the structures. The SDP approach turned out to increase the computational cost of the relaxed problem rapidly and required unreasonable time for the solution for any practical case. Continuous relaxation gave the worse approximation for all the test cases, compared to SQP and SA, but consumed less time to compute the answer. All materials needed for conducting an experiment can be found at \url{ https://sourceforge.net/p/mlalgorithms/code/HEAD/tree/Group374/Ryazanov2016InverseFolding/code/}.

Although the tested algorithms demonstrated a good capacity in solving the optimization problem, the actual quality of the primary structure prediction was  poor. This is mainly because the potential functions that we used for the energy estimation do not follow the assumption that the native sequence has the lowest energy for a given geometrical shape. Therefore, the structures that were found using the introduced techniques, had much lower energy than the native one, however the native and the predicted sequences coincided very weakly. We strongly believe that in order to achieve better prediction results, an improvement of the potential functions for the inverse folding problem is needed.\par

\section{Acknowledgments}
This work was supported by RFBR, grant \\16-37-00111.


\end{document}